\documentclass[prd,aps,a4paper,superscriptaddress,twocolumn,nofootinbib]{revtex4}
\usepackage{graphicx}
\usepackage{color}
\usepackage{dcolumn}
\usepackage{bm}
\usepackage{slashed}
\usepackage{amsmath}
\usepackage{latexsym}
\usepackage{amssymb}
\usepackage{mathrsfs}
\usepackage{amsfonts}
\usepackage{url}
\usepackage{graphicx}

\newcommand{\be}{\begin{equation}}
\newcommand{\ee}{\end{equation}}

\allowdisplaybreaks
\begin{document}
	
\title{Scale dependence of the effective gravitational constant from functional renormalization group}
	
\author{Ruiqi Liang}
\email[Ruiqi Liang:~]{liangruiqi24@mails.ucas.ac.cn}
\affiliation{School of Fundamental Physics and Mathematical Sciences, Hangzhou Institute for Advanced Study, University of Chinese Academy of Sciences, 
Hangzhou 310024, China}
\author{Zhoujian Cao\footnote{corresponding author}}
\email[Zhoujian Cao:~]{zjcao@bnu.edu.cn}
\affiliation{School of Fundamental Physics and Mathematical Sciences, Hangzhou Institute for Advanced Study, University of Chinese Academy of Sciences, 
Hangzhou 310024, China}
\affiliation{School of Physics and Astronomy, Beijing Normal University, Beijing 100875, China}
\author{Bing Sun}
\email[Bing Sun:~]{bingsun@mail.bnu.edu.cn}
\affiliation{Department of Basic Courses, Beijing University of Agriculture, Beijing 102206, China}

\begin{abstract}
The general relativity may not be the final theory of gravity. One possible way to look for the gravity theory beyond general relativity is considering variable gravitational constant. Based on the quantum field theory, the gravitational constant as the coupling constant of gravity interaction may change along with the energy scale. Such scale dependent behavior of the gravitational constant can be well described by the functional renormalization group. In the current work, we investigate such behavior systematically. Firstly we find that such behavior is qualitatively independent of interactions such as the electromagnetic interaction included or not. But quantitatively the behavior is governed by two to-be-determined parameters. In general a limit scale may be introduced by the scale dependence of the gravitational constant. If we assume all physical scales are feasible, the two parameters are limited in some special regions. And more we compare the scale dependence behavior to the existing observations. We find the observation results constraint the two parameters strictly.
\end{abstract}

\maketitle

\section{Introduction}
General relativity theory has developed more than one hundred years. Although it is still the most outstanding theory for gravity to explain all experiments to date, the singularity issue implies that it should fail in some physical conditions. Specially for understanding the singularity problem, many attentions are paid to the regular black hole model construction \cite{2023IJTP...62..202L,10.1088.1572-9494.ae662c}. If such regular black hole model is related to a diffeomorphism invariant theory, the model may be distinguished by observation \cite{6sws-hfj7}.

From the effective field theory viewpoint \cite{1995gr.qc....12024D,DUPUIS20211}, the interaction parameter may vary with scale or to say energy scale \cite{PhysRevD.57.971,PhysRevD.60.084011,PhysRevD.62.043008,2010JHEP...01..084D,Koch_2014,PhysRevD.111.064031}. That is to say the gravitational constant may behave as a function of scale instead of a constant \cite{RevModPhys.75.403,Uzan2011,An2023,SUN2024138350,AN2025102062}. It is interesting to investigate the modified behavior of black holes compared to the standard picture in general relativity.

As a fundamental interaction, electromagnetic interaction should always be there no matter the black hole is neutral or charged. This is because the quantum fluctuation of electromagnetic field is always there at least. So it is interesting to investigate the behavior of the effective gravitational constant when electromagnetic interaction is also included. More specifically we would like to check if the behavior of the effective gravitational constant will change when more interactions are considered.

The existing works \cite{PhysRevD.57.971,PhysRevD.62.043008,2010JHEP...01..084D,Koch_2014,PhysRevD.111.064031} considered the mutual effect between the gravitational constant and the cosmological constant. From the quantum gravity viewpoint, the cosmological constant should come from some more fundamental interactions \cite{2021PhLB..82336770Z}. That is to say these existing works have considered the effect of gravitational interaction and some interactions related to the cosmological constant on the scale dependence of the gravitational constant.

Compared to the existing works \cite{PhysRevD.57.971,PhysRevD.62.043008,2010JHEP...01..084D,Koch_2014,PhysRevD.111.064031}, in current paper we firstly retract one step to only consider gravitational interaction. Then we go one step forward to include the electromagnetic interaction. Through these studies we investigate the behavior of the scale dependence of the gravitational constant when different interactions are involved.

The arrangement of the rest of the current paper is as following. We deduce and explain the renormalization group equations when different interactions are involved in the next section. Then we analyze the behavior of the scale dependent gravitational constant through solving the renormalization group equations in Sec.~\ref{sec3}. In Sec.~\ref{sec4} we compare the scale dependence behavior of the gravitational constant to the existing observations. Finally we conclude the paper in the last section and some discussion is also presented there.

Throughout the paper we use natural units where $c=\hbar=1$. Within these units we have $G_N\approx6.7\times10^{-57}$eV${}^{-2}$. According to $G_N$ we have a reference energy scale $k_*$ which satisfies $k_*^2G_N=1$.
\section{The renormalization group equations}
Based on the dimension of the gravitational constant, the cosmological constant and the fine structure constant, we define the scale-dependent dimensionless quantities as
\begin{align}
&g_k\equiv k^2G_k,\\
&\lambda_k\equiv k^{-2}\Lambda_k,\\
&\alpha_k\equiv\alpha_k,
\end{align}
where $G_k$, $\Lambda_k$ and $\alpha_k$ are the scale-dependent gravitational constant, cosmological constant and fine structure constant respectively. Here $k$ is used to denote the scale or to say the typical momentum. UV (UltraViolet) region corresponds to $k\rightarrow\infty$ and IR (InfraRed) region corresponds to $k\rightarrow0$.

In this section we closely follow \cite{PhysRevD.57.971} and use the truncated the effective average action \cite{2010JHEP...01..084D} to derive the exact renormalization group equation for the gravitational constant. The effective average action is determined by a modified Legendre transform of the connected Green's function. Accordingly we can get the Wetterich equation \cite{1993PhLB..301...90W,Florchinger2010}. After that, the renormalization group equation can be deduced.

If we only consider gravitational interaction, we need only take $g_k$ into consideration. We introduce $t=\ln\frac{k}{k_*}$ which means the IR limit $k=0$ corresponds to $t=-\infty$ and the UV limit $k=\infty$ corresponds to $t=+\infty$. And the resulting renormalization group equation reads as
\begin{align}
\partial_t g_k &=\left(2 +\eta_N\right) g_k,\label{eq1}\\
\eta_N &= \frac{g_k B_1(0)}{1 - g_k B_2(0)},
\end{align}
where
\begin{align}
    B_1(y) &\equiv \frac{1}{12\pi} \Bigl[ 20\Phi^1_{1}(-2y) - 72\Phi^2_{2}(-2y) \Bigr. \nonumber \\
    &\Bigl. - 16\Phi^1_{1}(0) - 24\Phi^2_{2}(0) \Bigr], \\
    B_2(y) &\equiv -\frac{1}{24\pi} \Bigl[ 20\tilde{\Phi}^1_{1}(-2y) - 72\tilde{\Phi}^2_{2}(-2y) \Bigr],\\
\Phi_{n}^p(x)&\equiv\frac{1}{\Gamma(n)}\int_0^\infty dz z^3\frac{R^{(0)}(z)-zR^{(0)'}(z)}{[z+R^{(0)}(z)+x]^p},\\
\tilde{\Phi}_{n}^p(x)&\equiv\frac{1}{\Gamma(n)}\int_0^\infty dz z^3\frac{R^{(0)}(z)}{[z+R^{(0)}(z)+x]^p}.
\end{align}
The function $R^{(0)}(z)$ is called regulator involved in the renormalization group techniques. We can not completely determine the regulator function from first principle. The authors of \cite{PhysRevD.62.043008} choose the regulator function as
\begin{align}
R^{(0)}(z)=\frac{z}{e^z-1}.\label{eq9}
\end{align}
But definitely there are other possibilities for the regulator function choices. Based on the choice (\ref{eq9}), we have
\begin{align}
\eta_N &= \frac{144-\pi^2}{12g_k-18\pi}g_k.
\end{align}

When $g_k$ and $\lambda_k$ are considered which takes both the gravitational interaction and the cosmological constant involved interactions into consideration, we have the renormalization group equation \cite{PhysRevD.62.043008}
\begin{align}
&\partial_t g_k =\left(2 +\eta_N\right) g_k\label{eq2}\\
&\partial_t \lambda_k = -\left(2 -\eta_N\right) \lambda_k +  \nonumber\\
& \frac{g_k}{2\pi} \left[ 10\Phi_{2}^1 (-2\lambda_k) - 8 \Phi_{2}^1 (0) - 5\eta_N \tilde{\Phi}_{2}^1 (-2\lambda_k) \right].\label{eq4}
\end{align}
We can find out that Eq.~(\ref{eq1}) is exactly the same to Eq.~(\ref{eq2}). That is to say no matter we consider the cosmological constant involved interactions or not, the resulting scale dependence of $G$ is the same.

If both the gravitational interaction and the electromagnetic interaction are considered, both $g_k$ and $\alpha_k$ should be involved, and the resulting renormalization group equation reads as
\begin{align}
\partial_t g_k &= (2 + \eta_N)g_k,\label{eq3} \\
\partial_t \alpha_k &= \eta_F \alpha_k,\label{eq5}\\
\eta_F &\equiv -\frac{6 \Phi_1^1(0)}{\pi - 3 g_k \tilde{\Phi}_1^1(0)}g_k.
\end{align}
Again we find out that Eq.~(\ref{eq1}) is exactly the same to Eq.~(\ref{eq3}). That is to say the electromagnetic interaction does not affect the resulting scale dependence of $G$.

If we consider all of the gravitational interaction, the electromagnetic interaction and the cosmological constant involved interactions, $g_k$, $\alpha_k$ and $\lambda_k$ are coupled together as
\begin{align}
&\partial_t g_k = (2 + \eta_N)g_k, \label{eq6}\\
&\partial_t \alpha_k = \bar{\eta}_F \alpha_k, \label{eq7}\\
&\partial_t \lambda_k = -\left(2 -\eta_N\right) \lambda_k +  \nonumber\\
& \frac{g_k}{2\pi} \left[ 10\Phi_{2}^1 (-2\lambda_k) - 8 \Phi_{2}^1 (0) - 5\eta_N \tilde{\Phi}_{2}^1 (-2\lambda_k) \right],\label{eq8}\\
&\bar{\eta}_F \equiv -\frac{6 \Phi_1^1(0) +2 \eta_N \lambda_k}{\pi - 3 g_k \tilde{\Phi}_1^1(0) - 2 \lambda_k g_k}g_k.
\end{align}
The authors of \cite{2010JHEP...01..084D} have studied the renormalization group equation for coupling of the gravitational interaction, the Yang-Mills interaction and the cosmological constant involved interactions. If we specifically take the Yang-Mills field as an abelian field, the resulting renormalization group equation is the same as the above equation. At the mean time, we find that Eqs.(\ref{eq6}) and (\ref{eq8}) are the same to Eqs.~(\ref{eq2}) and (\ref{eq4}), and that Eqs.~(\ref{eq6}) and (\ref{eq7}) are the same to Eqs.~(\ref{eq3}) and (\ref{eq5}) with $\eta_F=\bar{\eta}_F(\lambda_k=0)$. That is to say, extra interactions do not affect the running coupling constant of the more fundamental interactions.

\section{Scale dependence of the effective gravitational constant}\label{sec3}
Based on the above section discussion, we can just investigate Eq.~(\ref{eq1}) to study the behavior of the scale dependence of the effective gravitational constant. Due to the uncertainty of the regulator function, there are two constants $B_{1,2}(0)$ which can not be determined from the first principle. In order to let our discussion more general we keep these two unknown constants and discuss the effect of these two constants.

We note that when $g_k$ approaches to $\frac{1}{B_2(0)}$, $\eta_N$ and consequently $\partial_tg_k$ diverge which should not happen for all scales if any scale is physically accessible. So if we assume all scales are physically accessible, we have $1-B_2(0)g_k(t)\neq0$. That is to say, as a continuous function of $t$, $1-B_2(0)g_k(t)$ can not change sign. Since $g_k(0)=1$, we have
\begin{align}
[1-B_2(0)][1-B_2(0)g_k(t)]>0,
\end{align}
which is equivalent to
\begin{align}
\text{if } &B_2(0)[1-B_2(0)] < 0 \nonumber\\
&g_k(t) > \frac{1}{B_2(0)}, \\
\text{if } &B_2(0)[1-B_2(0)] > 0 \nonumber\\
&g_k(t) < \frac{1}{B_2(0)}.
\end{align}
Combining with $g_k(0)=1$ we can rule out $B_2(0)=1$.

When $B_1(0)=2B_2(0)$, we can solve Eq.~(\ref{eq1}) as
\begin{align}
g_k=e^{2t+B_2(0)(g_k-1)}.
\end{align} But we find that if only $B_2(0)>0$
\begin{align}
&\frac{1}{B_2(0)}=e^{2t+B_2(0)(\frac{1}{B_2(0)}-1)},\\
&\frac{e^{B_2(0)}}{B_2(0)}=e^{1+2t}
\end{align}
always admits a solution of $t$. That is to say we must have $B_1(0)=2B_2(0)\leq0$, otherwise $g_k$ will approach to $\frac{1}{B_2(0)}$ at some $t$.

When $B_1(0)-2B_2(0)\neq0$ and $2+B_1(0)-2B_2(0)\neq0$, we can generally solve Eq.~(\ref{eq1}) as
\begin{align}
&Ce^{(2B_1(0)-4B_2(0))t}=\frac{1}{g_k^{2B_2(0)}}\times\nonumber\\
&\left(\frac{g_k}{2+(B_1(0)-2B_2(0))g_k}\right)^{B_1(0)},
\end{align}
where $C$ is the integral constant. According to the definition of $t$ we have $g_k(t=0)=1$, we can determine
\begin{align}
C=\left(\frac{1}{2+B_1(0)-2B_2(0)}\right)^{B_1(0)}.
\end{align}
So we have
\begin{align}
&e^{(2B_1(0)-4B_2(0))t}=\frac{1}{g_k^{2B_2(0)}}\times\nonumber\\
&\left(\frac{[2+B_1(0)-2B_2(0)]g_k}{2+[B_1(0)-2B_2(0)]g_k}\right)^{B_1(0)}.\label{eq15}
\end{align}

If at some $t_0$, $g_k=0$, Eq.~(\ref{eq1}) makes all $g_k$ vanish. Since $g_k(t=0)=1$, Eq.~(\ref{eq1}) guarantees that $g_k>0$. So the above solution requires
\begin{align}
&[2+B_1(0)-2B_2(0)][2+(B_1(0)-2B_2(0))g_k]>0.\label{eq10}
\end{align}

Since $B_1(0)-2B_2(0)\neq0$, the above requirement (\ref{eq10}) equivalents to
\begin{align}
\text{if } &[B_1(0) - 2B_2(0)] [2 + B_1(0) - 2B_2(0)] < 0 \nonumber\\
&g_k < \frac{2}{2B_2(0)-B_1(0)}, \label{eq11}\\
\text{if } &[B_1(0) - 2B_2(0)][2 + B_1(0) - 2B_2(0)] > 0 \nonumber\\
&g_k > \frac{2}{2B_2(0)-B_1(0)}. \label{eq18}
\end{align}
Then we have requirement for the two constants $B_{1,2}(0)$
\begin{align}
\text{if } &(B_1(0) - 2B_2(0)) [2 + B_1(0) - 2B_2(0)] < 0 \nonumber\\
&1 < \frac{2}{2B_2(0)-B_1(0)}, \\
\text{if } &(B_1(0) - 2B_2(0))[2 + B_1(0) - 2B_2(0)] > 0 \nonumber\\
&1 > \frac{2}{2B_2(0)-B_1(0)}.
\end{align}

Based on the assumption that all scales are physically available, $g_k$ can not approach to $\frac{1}{B_2(0)}$. Then combining (\ref{eq11}) and (\ref{eq18}) we have
\begin{align}
\text{if } &(B_1(0) - 2B_2(0)) [2 + B_1(0) - 2B_2(0)] < 0 \nonumber\\
&\frac{1}{B_2(0)}\geq \frac{2}{2B_2(0)-B_1(0)}, \\
\text{if } &(B_1(0) - 2B_2(0))[2 + B_1(0) - 2B_2(0)] > 0 \nonumber\\
&\frac{1}{B_2(0)}\leq\frac{2}{2B_2(0)-B_1(0)}.
\end{align}
which is equivalent to
\begin{align}
\text{if } &(B_1(0) - 2B_2(0)) [2 + B_1(0) - 2B_2(0)] < 0 \nonumber\\
&B_1(0)B_2(0)(B_1(0)-2B_2(0))\geq0, \\
\text{if } &(B_1(0) - 2B_2(0))[2 + B_1(0) - 2B_2(0)] > 0 \nonumber\\
&B_1(0)B_2(0)(B_1(0)-2B_2(0))\leq0.
\end{align}

In another word, if the above condition is not satisfied, some scale is physically unrealizable. The limit scale corresponds to $g_k=\frac{1}{B_2(0)}$.

\begin{figure}
\includegraphics[width=0.45\textwidth]{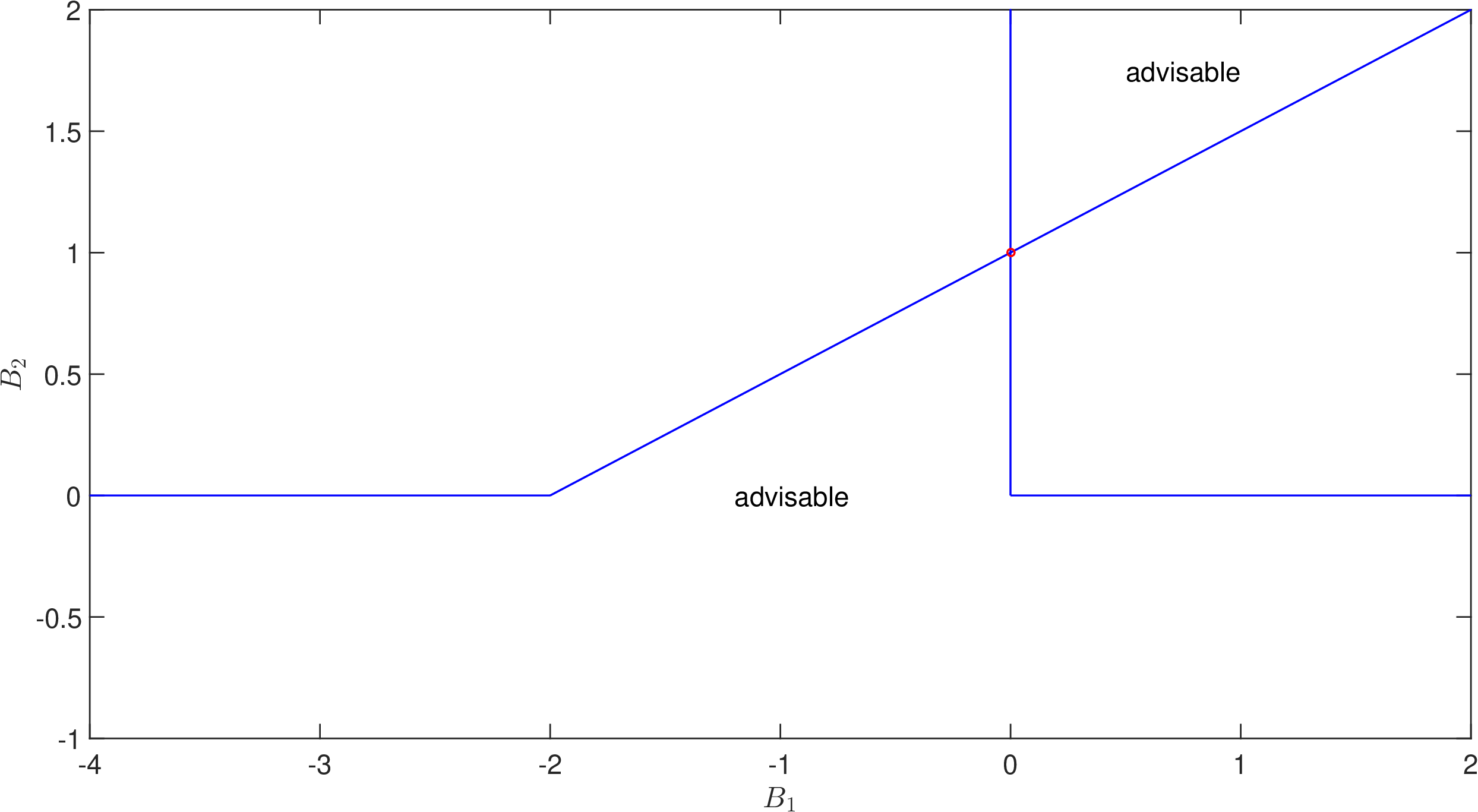}
\caption{The advisable regions (marked with `advisable') for parameters $B_1(0)$ and $B_2(0)$. The blue lines except the point marked with the red circle are advisable.}\label{fig0}
\end{figure}
When $B_2(0)\leq0$, since $g_k(t)$ is always positive, it will never approach to $1/B_2(0)$. Consequently if we assume all scales are accessible, we can summarize the requirement of the two constant $B_{1,2}(0)$ as following.
\begin{align}
&\text{if } B_2(0)\leq0, -\infty<B_1(0)<+\infty,\label{eq14}\\
&\text{if } 0<B_2(0)<1, 2(B_2(0)-1)<B_1(0)\leq0,\label{eq13}\\
&\text{if } B_2(0)\geq1, 0\leq B_1(0)<2(B_2(0)-1). \label{eq12}
\end{align}
Eq.~(\ref{eq12}) means $B_2(0)$ can not take the value 1. We summarize the advisable regions for $B_2-B_1$ plot in Fig.~\ref{fig0}.

\begin{figure}
\includegraphics[width=0.45\textwidth]{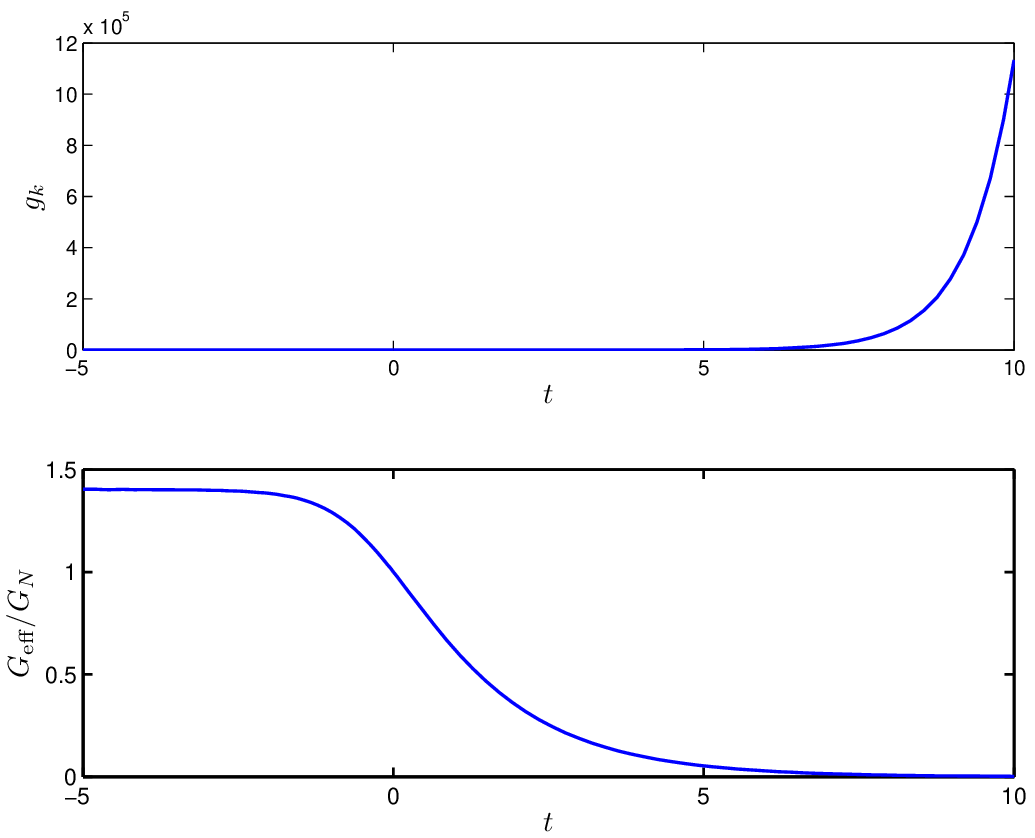}
\caption{Scale dependence of $g_k$ and the effective $G$ for case (\ref{eq14}) with specific values $B_1(0)=-1$ and $B_2(0)=-1.6$. Here $B_1(0)-2B_2(0)>0$.}\label{fig1}
\end{figure}

\begin{figure}
\includegraphics[width=0.45\textwidth]{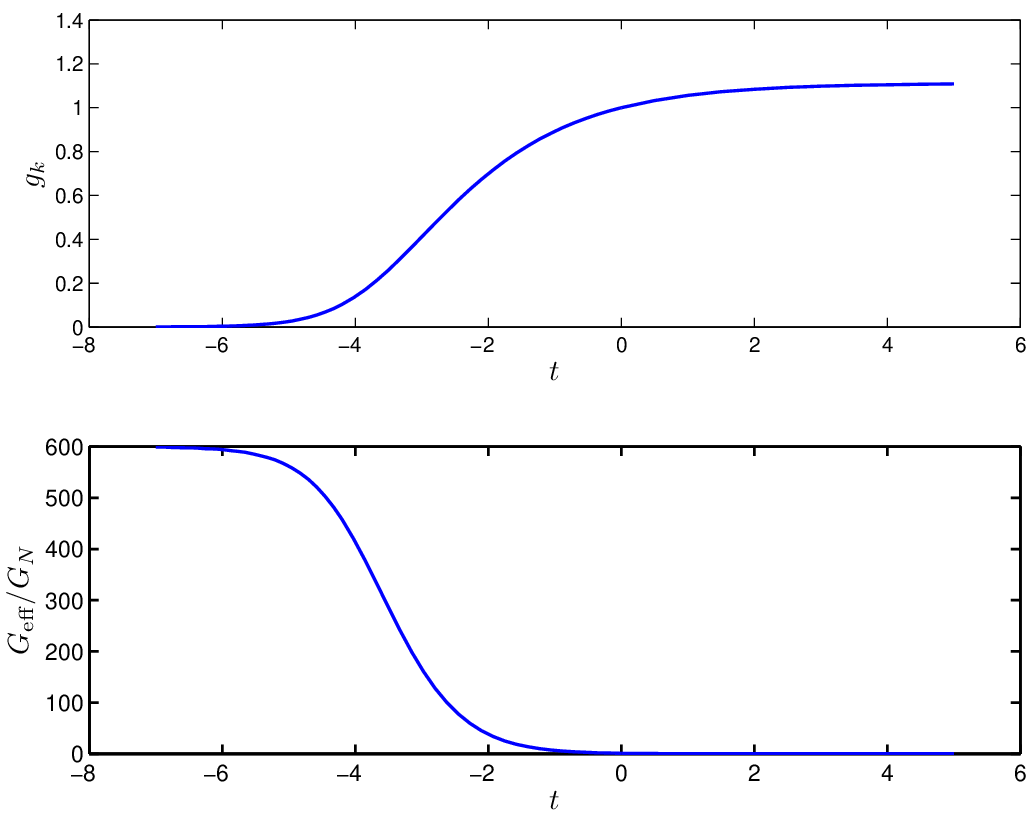}
\caption{Scale dependence of $g_k$ and the effective $G$ for case (\ref{eq14}) with specific values $B_1(0)=-5$ and $B_2(0)=-1.6$. Here $B_1(0)-2B_2(0)<0$.}\label{fig2}
\end{figure}
We take $B_1(0)=-1$, $B_2(0)=-1.6$ as the first example to investigate case (\ref{eq14}). We show the resulting effective $G$ in Fig.~\ref{fig1}. The UV limit of $G$ is zero which corresponds to asymptotical free and the Minkowski core in the regular black hole model. The IR limit of $G$ is about $1.404G_N$. The behavior of $G$ is similar to the one described in Fig.~2 of \cite{2023arXiv230204272P}. But the behavior of $g_k$ is different to that figure. We use $B_1(0)=-5$, $B_2(0)=-1.6$ as the second example to investigate this case. In the first example $B_1(0)-2B_2(0)>0$ which leads the UV limit of (\ref{eq15}) diverges. Instead in the second example this limit goes to zero. We show the resulted scale dependence in Fig.~\ref{fig2}. The IR limit of $G$ is very large. The behavior of $G$ and $g_k$ is similar to Fig.~2 of \cite{2023arXiv230204272P}.

\begin{figure}
\includegraphics[width=0.45\textwidth]{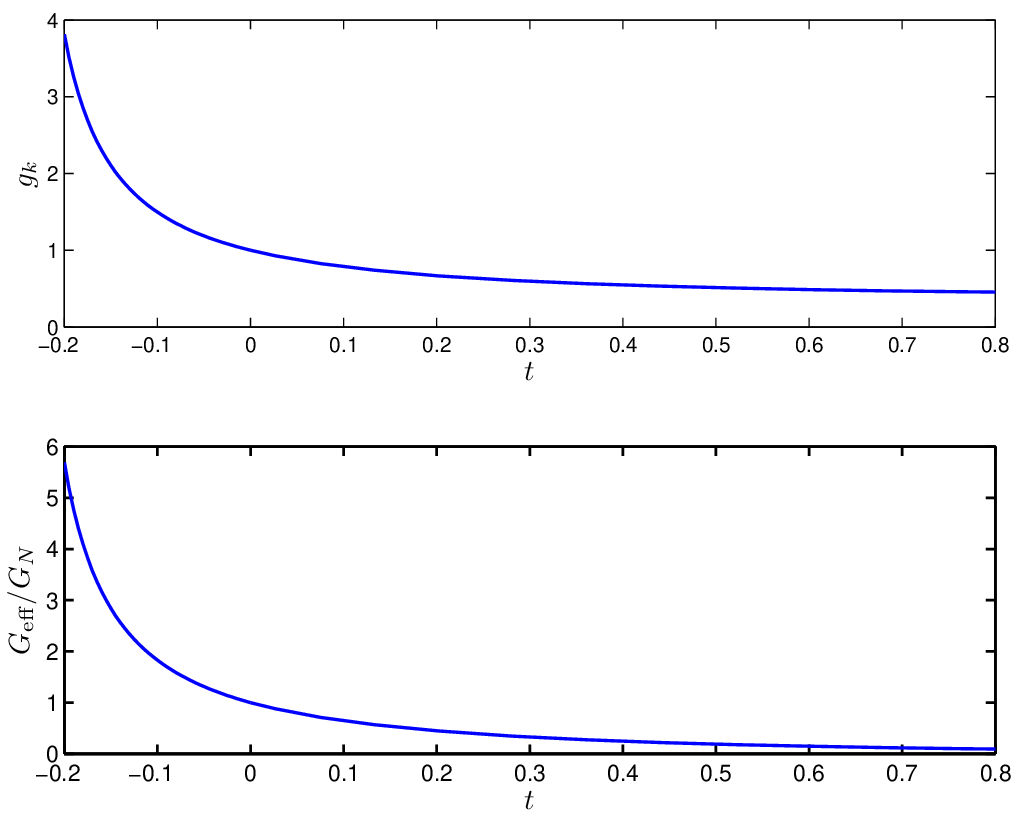}
\caption{Scale dependence of $g_k$ and the effective $G$ for parameters locating on the boundary of case (\ref{eq14}) with specific values $B_1(0)=-5$ and $B_2(0)=0$. Here $B_1(0)-2B_2(0)<0$.}\label{fig3}
\end{figure}

\begin{figure}
\includegraphics[width=0.45\textwidth]{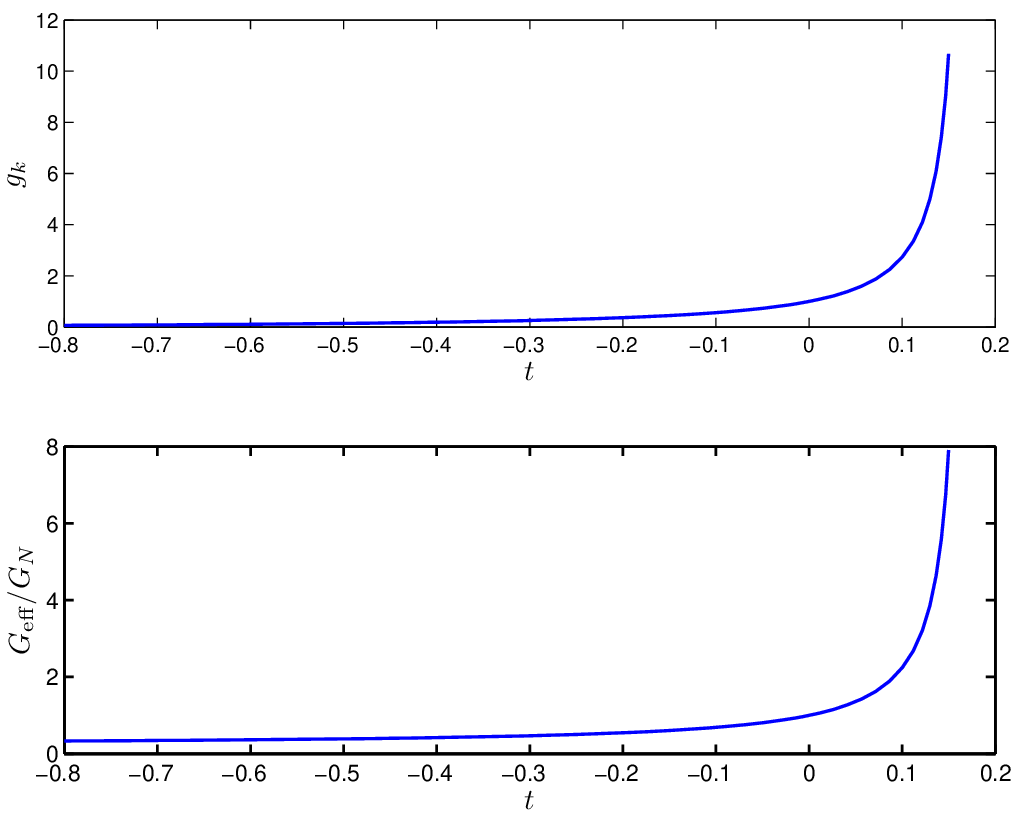}
\caption{Scale dependence of $g_k$ and the effective $G$ for parameters locating on the boundary of case (\ref{eq14}) with specific values $B_1(0)=5$ and $B_2(0)=0$. Here $B_1(0)-2B_2(0)>0$.}\label{fig4}
\end{figure}
For parameters locating on the boundary of case (\ref{eq14}) we firstly take $B_1(0)=-5$, $B_2(0)=0$ as the first example which corresponds to $B_1(0)-2B_2(0)<0$. The IR limit of both $G$ and $g_k$ diverge which is different to Fig.~2 of \cite{2023arXiv230204272P}. Corresponding to $B_1(0)-2B_2(0)>0$ we use $B_1(0)=5$, $B_2(0)=0$ as the second example for parameters locating on the boundary of case (\ref{eq14}). This example is completely different to Fig.~2 of \cite{2023arXiv230204272P}. Both $G$ and $g_k$ go to zero in IR region and diverge in UV limit which is shown in Fig.~\ref{fig4}.

\begin{figure}
\includegraphics[width=0.45\textwidth]{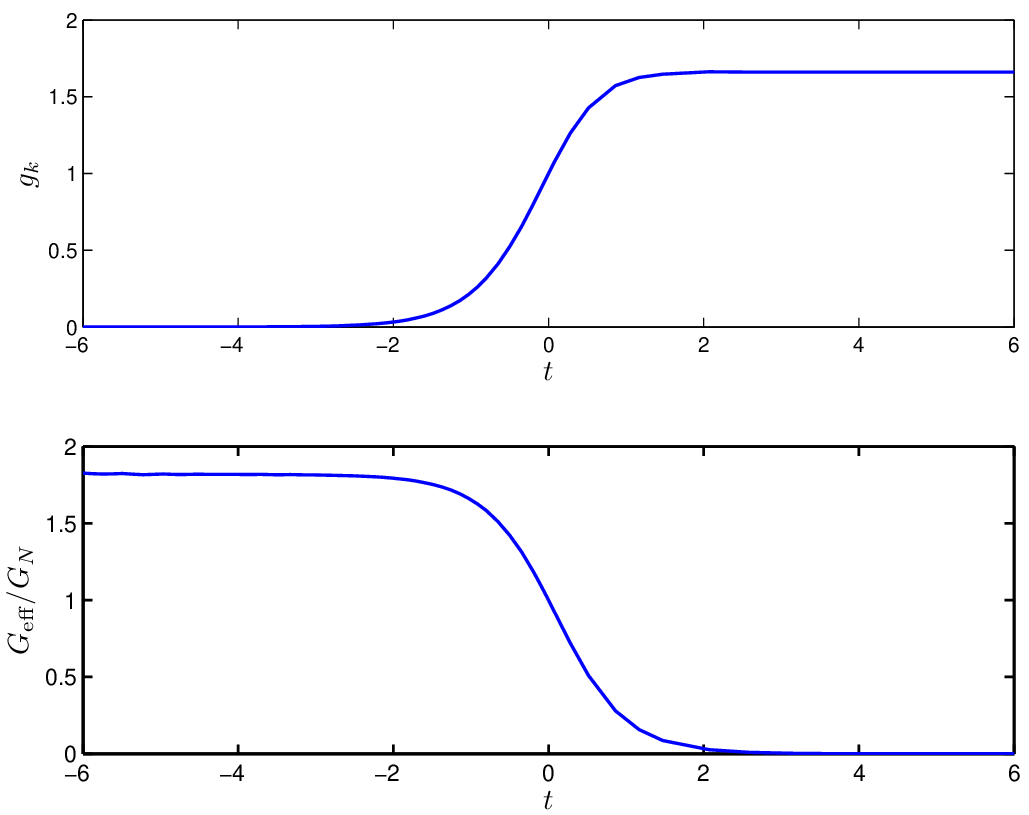}
\caption{Scale dependence of $g_k$ and the effective $G$ for regulator function (\ref{eq9}) which corresponds to case (\ref{eq13}) with specific values $B_1(0)\approx-0.78$ and $B_2(0)\approx0.21$. Here $B_1(0)-2B_2(0)<0$.}\label{fig5}
\end{figure}

For the regulator function (\ref{eq9}), we have
\begin{align}
B_1(0)&=\frac{\pi}{18}-\frac{8}{\pi},\\
B_2(0)&=\frac{2}{3\pi},
\end{align}
which falls in the case (\ref{eq13}). In this case the resulting $g_k$ and the effective $G$ are shown in Fig.~\ref{fig5}. The IR limit of $G$ is about $1.827G_N$. The qualitative behavior of $g_k$ and $G$ is similar to Fig.~\ref{fig2} and Fig.~2 of \cite{2023arXiv230204272P}. But the quantitative behavior is different. For the case (\ref{eq13}), $B_1(0)-2B_2(0)$ must be negative.

\begin{figure}
\includegraphics[width=0.45\textwidth]{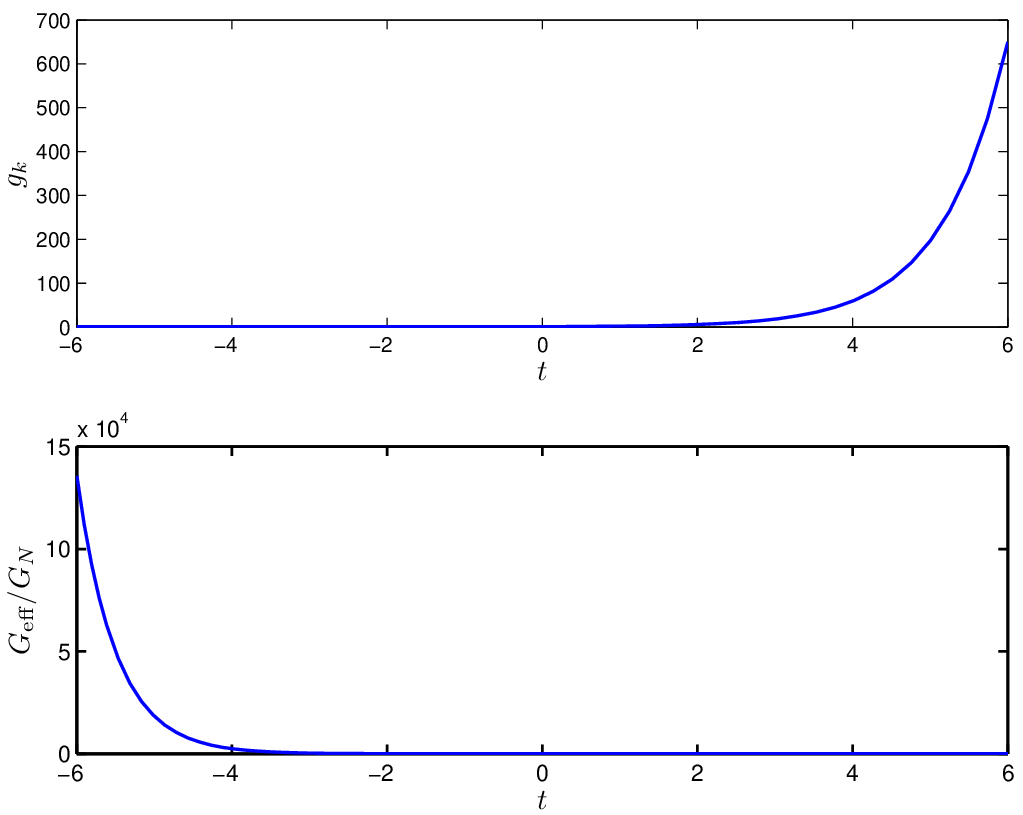}
\caption{Scale dependence of $g_k$ and the effective $G$ for case (\ref{eq12}) with specific values $B_1(0)=1.6$ and $B_2(0)=2$. Here $B_1(0)-2B_2(0)<0$.}\label{fig6}
\end{figure}
For the case (\ref{eq12}) we take $B_1(0)=1.6$, $B_2(0)=2$ as the example. The IR limit and UV limit of $g_k$ are both different to those of $G$. The IR limit of $G$ diverges and UV limit of $G$ goes to zero as shown in Fig.~\ref{fig6}. This behavior is different to Fig.~2 of \cite{2023arXiv230204272P}. For the case (\ref{eq12}) we have $B_1(0)-2B_2(0)<-2<0$.

\section{Observational constrain on the scale dependence of the effective gravitational constant}\label{sec4}

In order to relate the scale dependence behavior of the effective gravitational constant $G$ to real observation, we have to physically explain the meaning of scale $k$ or $t$ \cite{2025arXiv251014552S,Chen_2025}.

Using $G_k$, Eq.~(\ref{eq15}) can be equivalently expressed as
\begin{align}
&1=[2+B_1(0)-2B_2(0)]^{B_1(0)}\left(\frac{G_k}{G_N}\right)^{B_1(0)-2B_2(0)}\times\nonumber\\
&\left(\frac{1}{2+[B_1(0)-2B_2(0)]\frac{G_k}{G_N}e^{2t}}\right)^{B_1(0)}.\label{eq16}
\end{align}
Specifically if $B_1(0)=0$, $G_k=G_N$ is scale independent.

If observations constraint $aG_N<G_k<bG_N$ at some scale $t_0$, we can accordingly constraint parameters $B_{1,2}(0)$. As an example, if we consider $B_2(0)=0$ Eq.~(\ref{eq16}) becomes
\begin{align}
B_1(0)=\frac{2}{e^{2t_0}-1}\left(1-\frac{G_N}{G_k}\right).
\end{align}
Then we can constraint
\begin{align}
\frac{2}{e^{2t_0}-1}\left(1-\frac{1}{a}\right)<B_1(0)<\frac{2}{e^{2t_0}-1}\left(1-\frac{1}{b}\right).\label{eq19}
\end{align}

If we denote $\alpha\equiv\frac{G_k}{G_N}$, $x\equiv B_1(0)-2B_2(0)$ and $y\equiv B_1(0)$, Eq.~(\ref{eq16}) can be simplified as
\begin{align}
y=-\frac{x\ln\alpha}{\ln[(2+x)/(2+\alpha x e^{2t})]}.
\end{align}
When
\begin{align}
\min(-2,-\frac{2}{\alpha e^{2t}})<x<\max(-2,-\frac{2}{\alpha e^{2t}}),\label{eq20}
\end{align}
$(2+x)/(2+\alpha x e^{2t})<0$ and consequently the above $y$ does not exist.

The observations constraint $aG_N<G_k<bG_N$ is equivalent to $a<\alpha<b$ which means for a given $x$ we have
\begin{align}
&y_{\rm min}<y<y_{\rm max},\\
&y_a\equiv-\frac{x\ln a}{\ln[(2+x)/(2+a x e^{2t_0})]},\\
&y_b\equiv-\frac{x\ln b}{\ln[(2+x)/(2+b x e^{2t_0})]},\\
&y_{\rm min}\equiv \min(y_a,y_b),\\
&y_{\rm max}\equiv \max(y_a,y_b).
\end{align}
At the mean time, the condition (\ref{eq20}) gives requirement $x<\min(-2,-\frac{2}{a e^{2t_0}},-\frac{2}{b e^{2t_0}})$ or $x>\max(-2,-\frac{2}{a e^{2t_0}},-\frac{2}{b e^{2t_0}})$. Then we can accordingly get the permitted parameters region for $B_1(0)-B_2(0)$ space.
\subsection{Relate the scale $k$ to distance}

\begin{figure}
\includegraphics[width=0.45\textwidth]{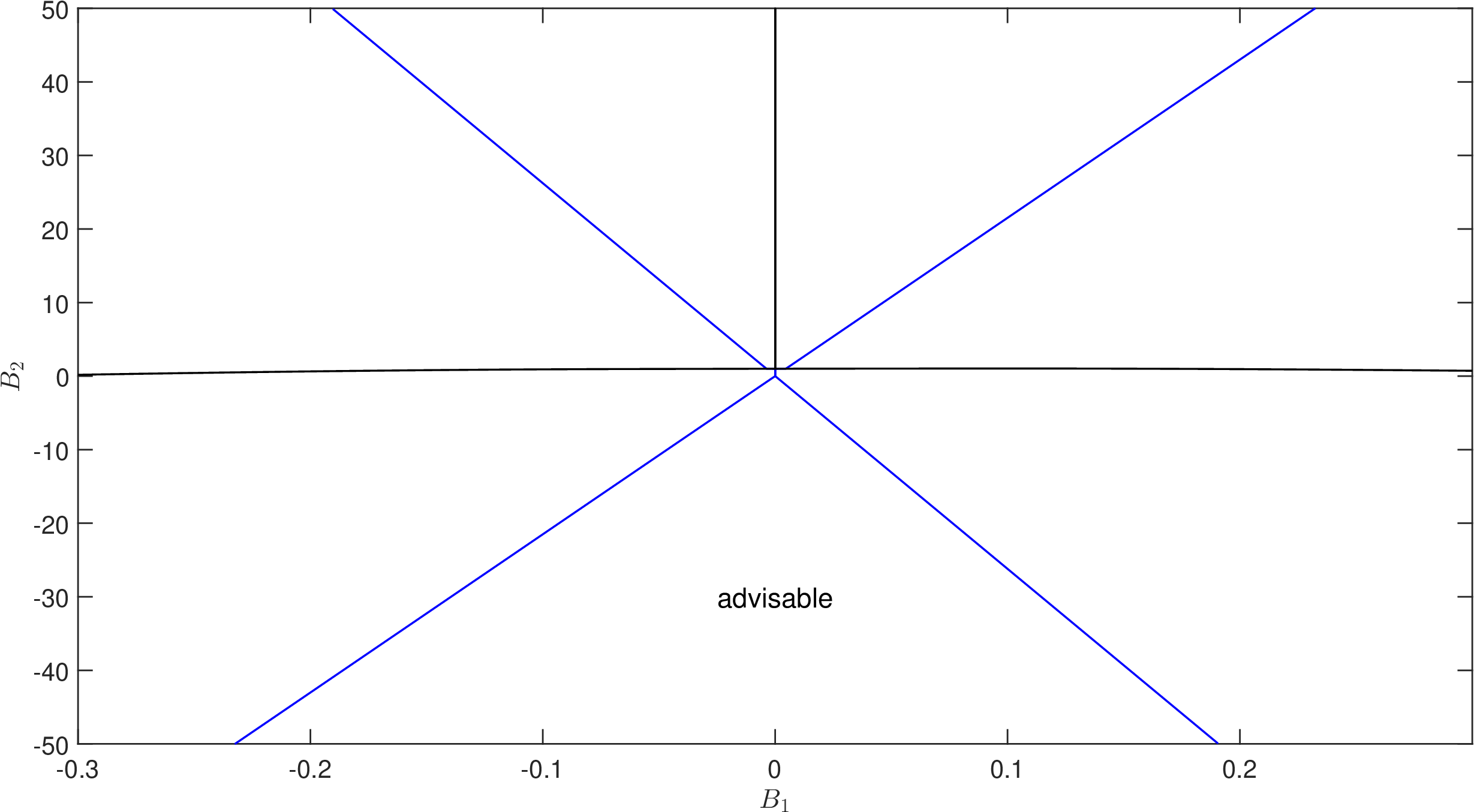}\\
\vspace{0.4cm}
\includegraphics[width=0.45\textwidth]{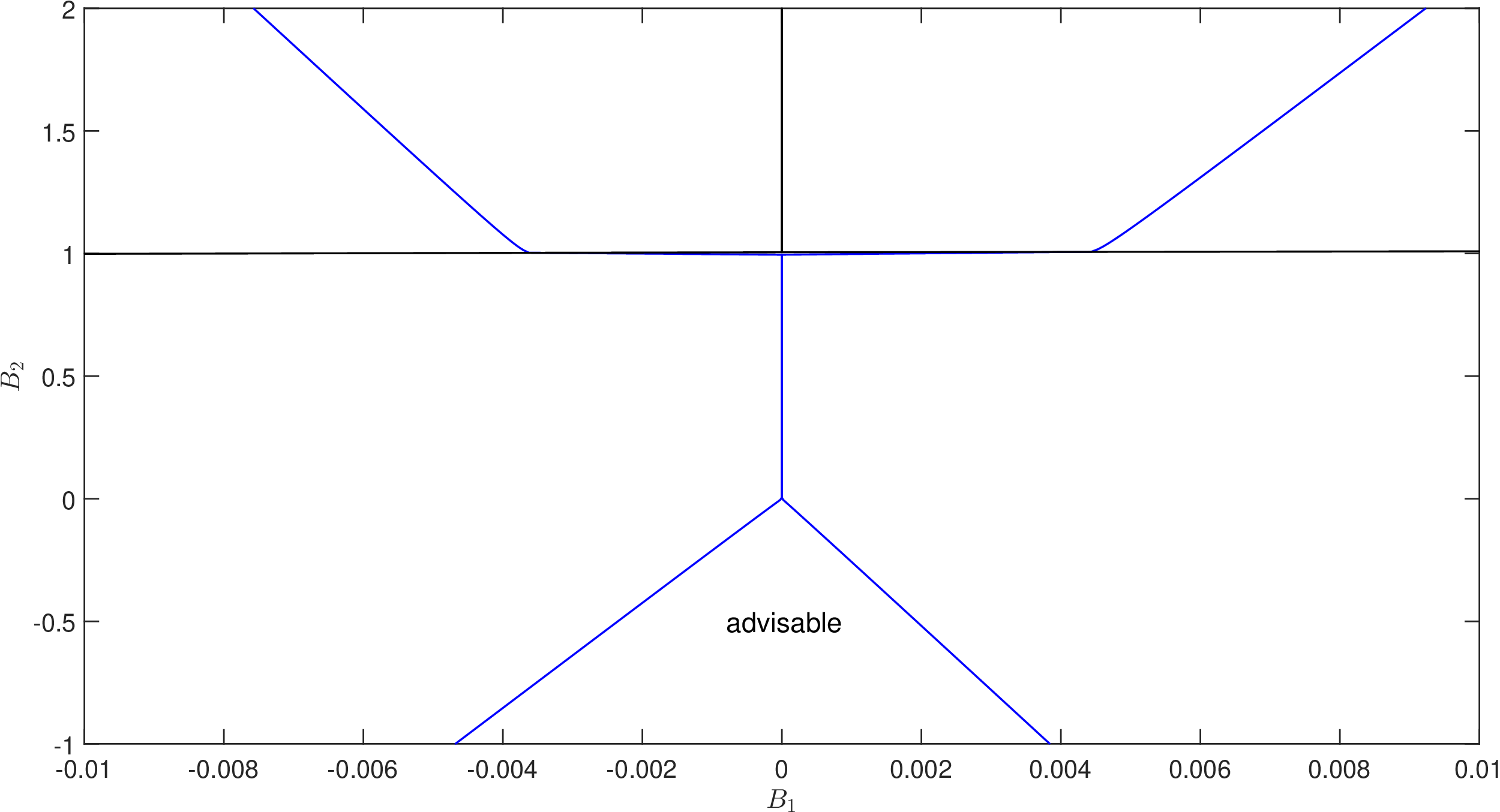}
\caption{Observation constraint of parameters $B_1(0)$ and $B_2(0)$ based on GW170817 by taking $k$ as distance. The blue lines correspond to the constraint of $t=48$ and the black lines correspond to $t=-48$. The bottom subplot is the zooming in around the region around original point which corresponds to Fig.~\ref{fig0}.}\label{fig8}
\end{figure}

One possibility of the scale $k$ is the inverse of distance or length \cite{PhysRevD.62.043008,PhysRevD.111.064031}. Current observations from cm scale to $10^{14}$m scale indicate that the Newtonian square inverse law is valid \cite{2009PrPNP..62..102A}. But it is hard to constrain the position dependence the effective $G$. Using gravitational wave observations, GW170817 has constrain the effective $G$ difference between the source and the observer (about 40Mpc separation) is less than 20\% ($\left|\frac{\Delta G}{G_N}\right|\lesssim0.2$) \cite{An2023,SUN2024138350,AN2025102062}. That is to say the binary neutron star with separation about km feels roughly the same effective $G$ as the binary neutron star and the observer with separation about 40Mpc. The scale difference between km and 40Mpc is about $e^t\thickapprox1.2\times10^{21}$. This means the gravitational wave observation constrain $0.8G_N\lesssim G_k(t=-48),G_k(t=48)\lesssim1.2G_N$. This constrain is very stringent. All the parameters setting from Fig.~\ref{fig1}-\ref{fig6} are ruled out. We plot the constraint results in Fig.~\ref{fig8}. In the center of the plot the advisable region around $B_1(0)=0$ behaves as almost a line. From (\ref{eq19}) we can estimate the advisable region is about $-10^{-42}\lesssim B_1(0)\lesssim10^{-42}$.

\begin{figure}
\includegraphics[width=0.45\textwidth]{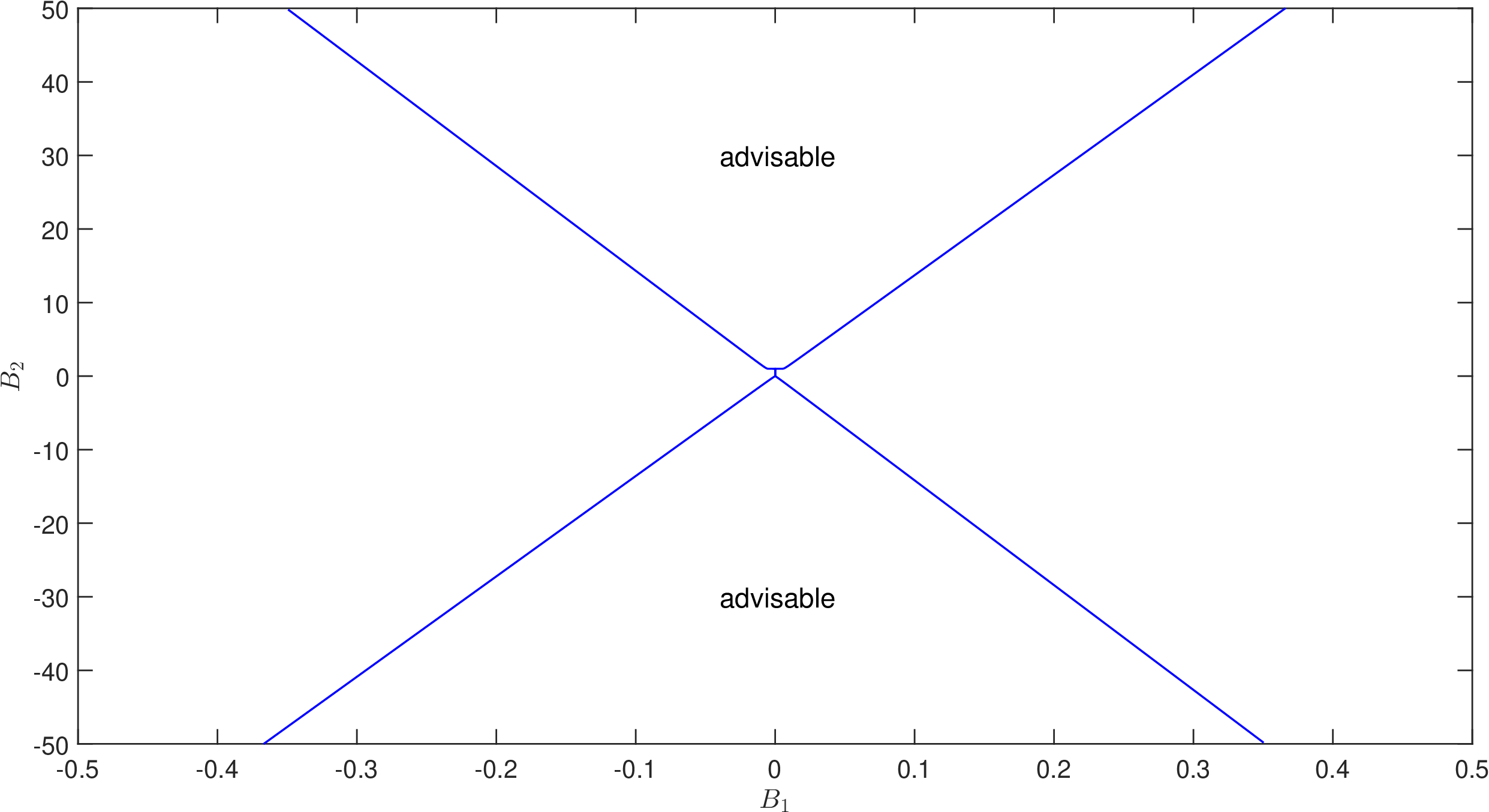}\\
\vspace{0.4cm}
\includegraphics[width=0.45\textwidth]{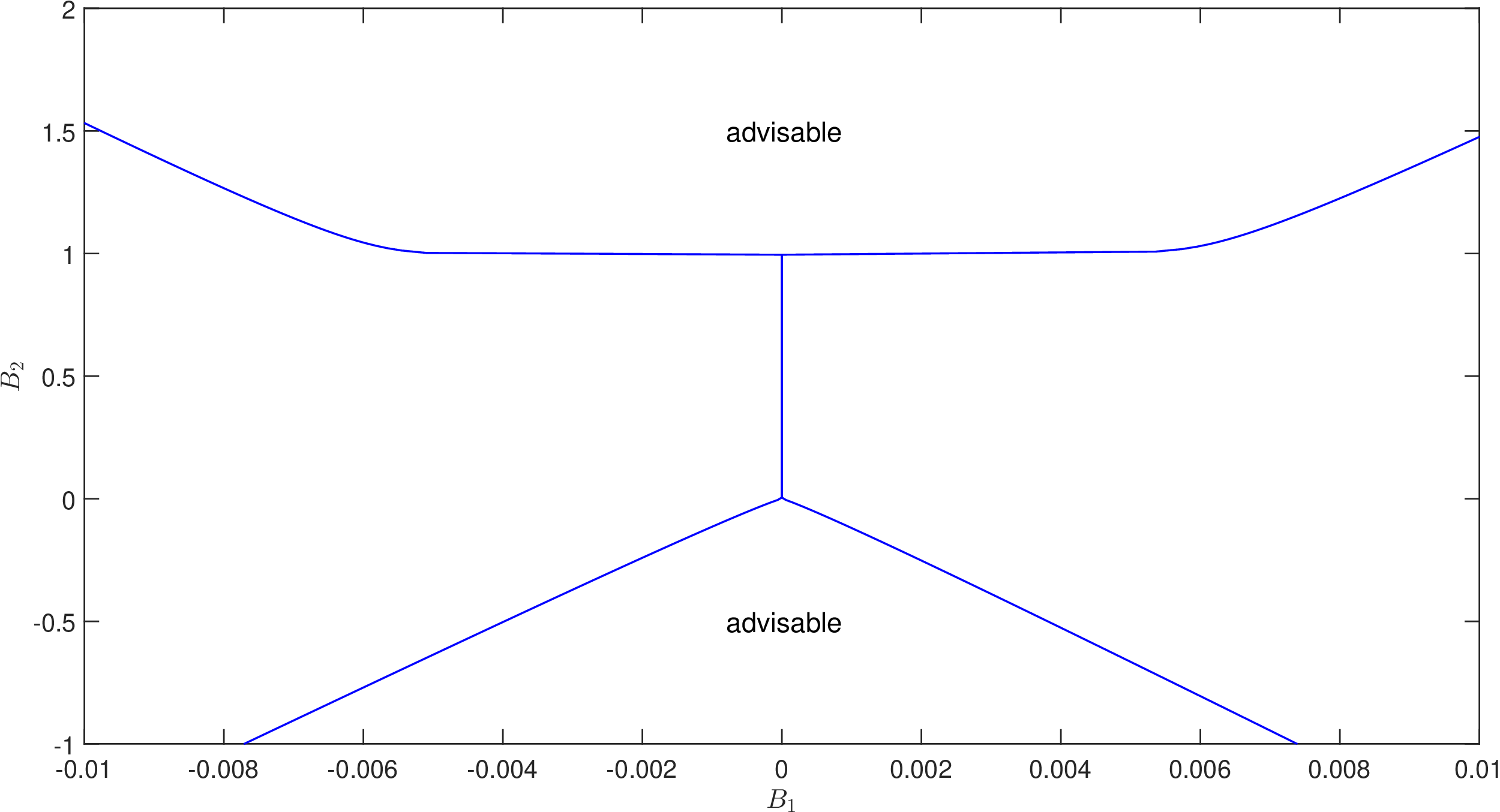}
\caption{Observation constraint of parameters $B_1(0)$ and $B_2(0)$ based on cosmology by taking $k$ as distance. The bottom subplot is the zooming in around the region around original point which corresponds to Fig.~\ref{fig0}.}\label{fig9}
\end{figure}

Reminding cosmology, we can relate $k$ to the cosmological scale factor $a$ \cite{PhysRevD.65.043508}. The observations indicate that the effective $G$ at the recombination epoch ($a\thickapprox1/1100$) is roughly the same to the current ($a=1$) effective $G$, $\left|\frac{\Delta G}{G_N}\right|\lesssim0.05$ \cite{PhysRevD.69.083512}. This means $0.95G_N\lesssim G_k(t=7)\lesssim1.05G_N$. This constrain is also very stringent. All the parameters setting from Fig.~\ref{fig1}-\ref{fig6} are ruled out. We plot the constraint results in Fig.~\ref{fig9}. In the center of the plot the advisable region around $B_1(0)=0$ behaves as almost a line. From (\ref{eq19}) we can estimate the advisable region is about $-10^{-8}\lesssim B_1(0)\lesssim10^{-8}$.
\subsection{Relate the scale $k$ to time $\tau$}

Current observations constrain the time variation of $G$ to less than $10^{-14}$yr${}^{-1}$ ($\left|\frac{\dot{G}}{G_N}\right|<10^{-14}$yr${}^{-1}$) \cite{2009PrPNP..62..102A}. According to (\ref{eq1}) and relating the scale $k$ to time $k=1/\tau$ \cite{PhysRevD.65.043508} we have
\begin{align}
&\left.\frac{dG/dk}{G_N}\right|_{k=k_*}=\frac{\eta_N}{k_*},\\
&\left.\frac{dG/d\tau}{G_N}\right|_{k=k_*}=-k_*\eta_N.
\end{align}
Since $k_*\approx1.83\times10^{43}$s${}^{-1}\approx5.76\times10^{50}$yr${}^{-1}$ the observation requires
\begin{align}
&\left|\eta_N\right|\lesssim1.7 \times 10^{-65},\label{eq17}\\
&\left|\frac{B_1(0)}{1-B_2(0)}\right|\lesssim1.7 \times 10^{-65}.
\end{align}
This constraint is very strict. It means either $\left|B_1(0)\right|$ is very small $\left|B_1(0)\right|\lesssim10^{-65}$ or $\left|B_2(0)\right|$ is very big $\left|B_2(0)\right|\gtrsim10^{65}$. All parameters used in the examples of Sec.~\ref{sec3} do not satisfy this requirement.

The lunar laser ranging has also constrain $\left|\frac{\ddot{G}}{G_N}\right|\lesssim10^{-15}$yr${}^{-2}$ \cite{Muller_2007,Merkowitz2010}. Since
\begin{align}
&\left.\frac{d^2G/d\tau^2}{G_N}\right|_{k=k_*}=k^2_*\eta_N(1+\eta_N),
\end{align}
we have
\begin{align}
&\left|\eta_N(1+\eta_N)\right|\lesssim3\times10^{-117},\\
&\left|\eta_N\right|\lesssim3\times10^{-117}.
\end{align}
This constraint is even more strict than (\ref{eq17}).

For cosmology there is no reasonable reference time point to use observation to constrain the variation of $G$. But for gravitational wave, the emitting time is the natural reference time point. For GW170817 we can use the gravitational wave period to estimate the time at source side $\tau\approx10$ms. Propagating to the earth, the time lasts $\tau\approx4\times10^{15}$s. So the gravitational wave observation \cite{An2023,SUN2024138350,AN2025102062} leads to $0.8G_N\lesssim G_k(t\approx\pm40.5)\lesssim1.2G_N$. This constrain is also very stringent. All the parameters setting from Fig.~\ref{fig1}-\ref{fig6} are ruled out. The constraint is similar to that of Fig.~\ref{fig8} but a little bit looser. The advisable region around the original point is about $-10^{-36}\lesssim B_1(0)\lesssim10^{-36}$.

\subsection{Relate the scale $k$ to spacetime curvature}
One may relate the scale $k$ to the Kretschmann scalar \cite{2019JCAP...06..029H,2025arXiv251014552S}. In principle the black hole shadow observation can check the scale dependence of the gravitational constant \cite{2019JCAP...06..029H}. But the requirement of the instrument accuracy is very high.

We can again use cosmological observation to do the constrain. For the Friedmann-Robertson-Walker metric, the Kretschmann scalar reads
\begin{align}
K=12\left[\left(\frac{\dot{a}}{a}\right)^4+\left(\frac{\ddot{a}}{a}\right)^2\right].
\end{align}
Currently we have $a=1$, $\frac{\dot{a}}{a}=H_0$, $\frac{\ddot{a}}{a}=-q_0H^2_0$ where $H_0\approx70$km/s/Mpc$\approx2\times10^{-18}$s${}^{-1}$ and $q_0\approx-0.4$ are the Hubble constant and the cosmological deceleration parameter \cite{Wambsganss_1997,aghanim2020planck6}. So we have $K_*\approx2.2\times10^{-70}$s${}^{-4}$ at present time. Relating this $K_*$ to $k_*\approx1.83\times10^{43}$s${}^{-1}$ we get
\begin{align}
k\approx4.7\times10^{60}K^{1/4}.
\end{align}

At the recombination epoch we have $a\approx1/1100$, and the standard cosmological model with cold dark matter and cosmological constant ($\Lambda$CDM) predicts $\frac{\dot{a}}{a}=H\approx10^5$km/s/Mpc$\approx10^{-15}$s${}^{-1}$ and $q\equiv-\frac{\ddot{a}a}{\dot{a}^2}\approx1$ which leads $\frac{\ddot{a}}{a}=-qH^2\approx10^{-30}$s${}^{-2}$ . Consequently we have $K\approx10^{-59}$s${}^{-4}$ at the recombination epoch. So the observation constrain \cite{PhysRevD.69.083512} leads $0.95G_N\lesssim G_k(t\approx3.2)\lesssim1.05G_N$. This constrain is also very stringent. All the parameters setting from Fig.~\ref{fig1}-\ref{fig6} are ruled out. The constraint is similar to that of Fig.~\ref{fig9} but a little bit looser. The advisable region around the original point is about $-10^{-4}\lesssim B_1(0)\lesssim10^{-4}$.

Regarding the gravitational wave observation \cite{An2023,SUN2024138350,AN2025102062}, we can use Schwarzschild metric to estimate the Kretschmann scalar
\begin{align}
K=48G_N^2\frac{M^2}{r^6}
\end{align}
for the site of GW170817 gravitational wave source and the site of observer. For the source site, $M\approx2M_\odot$ and $r\approx10$km (we use the wave zone to estimate the source site). For the observation site, the gravitational mass corresponds to the earth $M\approx3\times10^{-6}M_\odot$ and $r\approx6371$km. So the observation constrain leads to $0.8G_N\lesssim G_k(t\approx16.4)\lesssim1.2G_N$. This constrain is also very stringent. All the parameters setting from Fig.~\ref{fig1}-\ref{fig6} are ruled out. The constraint is similar to that of Fig.~\ref{fig9} but a little bit stricter. The advisable region around the original point is about $-10^{-15}\lesssim B_1(0)\lesssim10^{-15}$.

\subsection{Relate the scale $k$ to temperature}

Scale $k$ has meaning of energy. Since temperature $T$ corresponds to energy naturally, it is possible relating the scale $k$ to temperature
\begin{align}
k\propto k_BT,
\end{align}
where $k_B$ is the Boltzmann constant. Once again for the cosmological observation, we have current temperature $T\approx2.7$K and the temperature at the recombination epoch $T\approx3000$K. So the observation constrain \cite{PhysRevD.69.083512} leads $0.95G_N\lesssim G_k(t\approx7)\lesssim1.05G_N$. This constrain is also very stringent. All the parameters setting from Fig.~\ref{fig1}-\ref{fig6} are ruled out. The constraint is similar to that of Fig.~\ref{fig9}. The advisable region around the original point is also about $-10^{-8}\lesssim B_1(0)\lesssim10^{-8}$.

\section{Conclusion and discussion}
According to the viewpoint of quantum field theory, the coupling constant may change along with energy scales. Taking the gravity as the fourth interaction form, the Newtonian gravitational constant is the corresponding coupling constant between the gravity charge, mass energy, and the gravity field, spacetime metric. So it is natural to argue that the effective gravitational constant may also change with the energy scales.

The functional renormalization group method is a powerful tool to study the scale dependence of coupling constant. Closely follow this method we find out that no matter we consider other interactions than gravity or not, the government equation of the scale dependence of the effective gravitational constant does not change. At least the form does not change. It is always described by Eq.~(\ref{eq1}). But there are two undetermined parameters $B_1(0)$ and $B_2(0)$ involved. These two parameters is in principle determined by the regulator function in the functional renormalization group. This regulator function may be affected by the interactions being considered. But anyhow the equation form always takes Eq.~(\ref{eq1}).

We take an agnostic viewpoint to analyze Eq.~(\ref{eq1}). By investigating its solution, we find out that a limit scale exists for some parameters choices of $B_{1,2}(0)$. If we assume all scales are feasible, only specific regions of $B_1(0)$-$B_2(0)$ parameter space are advisable. The advisable regions are summarized in Fig.~\ref{fig0}.

The solution of Eq.~(\ref{eq1}) presents us the scale dependence behavior of the effective gravitational constant. Comparing this solution to the observation results, we can give a constraint to the parameters $B_1(0)$ and $B_2(0)$. We find the constraint is very strict. The constraint result is roughly either $B_2(0)<0$ or $|B_1(0)|<10^{-42}$.

\acknowledgments
This work is supported by the National Key Research and Development Program of China (Nos. 2021YFC2203001), the National Natural Science Foundation of China (No.~12475049).

\bibliographystyle{unsrt}
\bibliography{refs}

\end{document}